\documentclass[twocolumn,showpacs,preprintnumbers]{revtex4}%
\usepackage{amsfonts}
\usepackage{amsmath}
\usepackage{amssymb}
\usepackage{graphicx}%
\providecommand{\U}[1]{\protect\rule{.1in}{.1in}}
\providecommand{\U}[1]{\protect\rule{.1in}{.1in}}
\providecommand{\U}[1]{\protect\rule{.1in}{.1in}}
\providecommand{\U}[1]{\protect\rule{.1in}{.1in}}
\providecommand{\U}[1]{\protect\rule{.1in}{.1in}}
\providecommand{\U}[1]{\protect\rule{.1in}{.1in}}
\providecommand{\U}[1]{\protect\rule{.1in}{.1in}}

\begin{document}
\preprint{ }
\title{Detection of Fermi Pairing via Electromagnetically Induced Transparency}
\author{Lei Jiang$^{1}$, Han Pu$^{1}$, Weiping Zhang$^{2}$, and Hong Y. Ling$^{3}$}
\author{}
\affiliation{$^{1}$Department of Physics and Astronomy, and Rice Quantum Institute, Rice
University, Houston, TX 77251, USA }
\affiliation{$^{2}$State Key Laboratory of Precision Spectroscopy, Department of Physics,
East China Normal University, Shanghai 200062, P. R. China }
\affiliation{$^{3}$Department of Physics and Astronomy, Rowan University, Glassboro, New
Jersey, 08028-1700, USA}

\begin{abstract}
An optical spectroscopic method based on the principle of
electromagnetically-induced transparency (EIT) is proposed as quite
a generic probing tool that provides valuable insights into the
nature of Fermi paring in ultracold Fermi gases of two hyperfine
states. This technique has the capability of allowing spectroscopic
response to be determined in a nearly non-destructive manner and the
whole spectrum may be obtained by scanning the probe laser frequency
faster than the lifetime of the sample without re-preparing the
atomic sample repeatedly. A quasiparticle picture is constructed to
facilitate a simple physical explanation of the pairing signature in
the EIT spectra.

\end{abstract}
\date{\today }

\pacs{03.75.Ss, 05.30.Fk, 32.80.Qk}
\maketitle

\section{Introduction}

The two-component degenerate Fermi gas, in which the interaction between atoms
of two different hyperfine states is made magnetically tunable via Feshbach
resonance, has been the main source of inspiration for much recent excitement
at the forefront of ultracold atomic physics research. In addition to being an
ideal system for the exploration of the crossover from a Bose-Einsten
condensate (BEC) of highly localized pairs to nonlocal
Bardeen-Cooper-Schrieffer (BCS) pairs, the degenerate Fermi gas, when
operating in the unitarity regime, constitutes a strongly interacting Fermi
gas exhibiting a rich set of physics, the study of which may shed light on
long-standing problems in many different branches of physics, in particular,
condensed matter physics.

A unique phenomenon of low temperature Femi system is the formation
of correlated Fermi pairs. How to detect pair formation in an
indisputable fashion has remained a central problem in the study of
ultracold atomic physics. Unlike the BEC transition of bosons for
which the phase transition is accompanied by an easily detectable
drastic change in atomic density profile, the onset of pairing in
Fermi gases does not result in measurable changes in fermion
density. Early proposals sought the BCS pairing signature from the
images of off-resonance scattering light \cite{zhang99}. The
underlying idea is that to gain pairing information, measurement
must go beyond the first-order coherence, for example, to the
density-density correlation. This is also the foundation for other
detecting methods such as spatial noise correlations in the image of
the expanding gas \cite{altman04}, Bragg scattering
\cite{minguzzi01,bragg}, Raman spectroscopy \cite{torma00}, Stokes
scattering method \cite{baym}, radio frequency (RF) spectroscopy
\cite{chin04,torma04}, optical detection of absorption
\cite{ruostekoski99}, and interferometric method \cite{castin05}.
Among all these methods, RF spectroscopy \cite{chin04,torma04} has
been the only one implemented in current experiments
\cite{Regal03,Gupta03}.

\begin{figure}
[ptb]
\begin{center}
\includegraphics[
height=3.1415in,
width=3.0676in
]%
{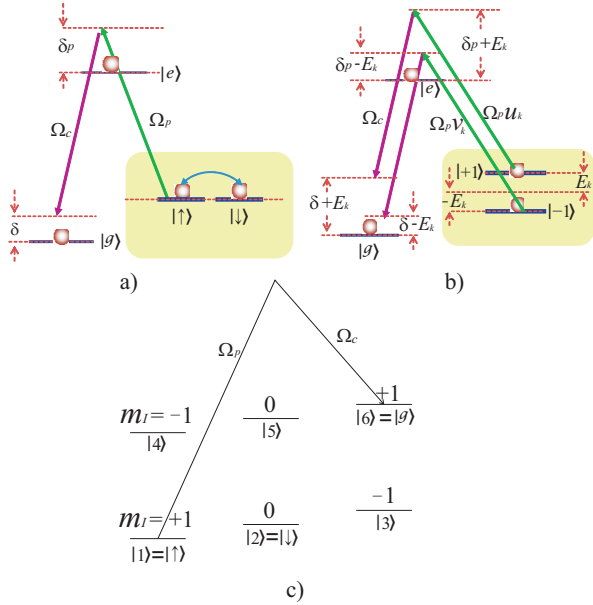}%
\caption{(Color online) (a) The bare state picture of our model. (b) The
dressed state picture of our model equivalent to (a). (c) A possible
realization in $^{6}$Li. Here the states labelled by $\left\vert
i\right\rangle $ ($i=1,2,...,6$) are the $6$ ground state hyperfine states.
Most experiments involving $^{6}$Li are performed with a magnetic field
strength tuned near a Feshbach resonance at 834G. Under such a magnetic field,
the magnetic quantum number for the nuclear spin $m_{I}$ is, to a very good
approximation, a good quantum number. The values of $m_{I}$ are shown in the
level diagrams. Two-photon transition can only occur between states with the
same $m_{I}$. Any pair of the lower manifold ($\left\vert 1\right\rangle $,
$\left\vert 2\right\rangle $, and $\left\vert 3\right\rangle $) can be chosen
to form the pairing states. In the example shown here, we choose $\left\vert
1\right\rangle =\left\vert \uparrow\right\rangle $, $\left\vert 2\right\rangle
=\left\vert \downarrow\right\rangle $ and $\left\vert 6\right\rangle
=\left\vert g\right\rangle $. The excited state $\left\vert e\right\rangle $
(not shown) can be chosen properly as one of the electronic $p$ state.}%
\label{Fig:schematics}%
\end{center}
\end{figure}

In this paper, we propose an alternative detection scheme, whose
principle of operation is illustrated in
Fig.~\ref{Fig:schematics}(a). In our scheme, a relatively strong
coupling and a weak probe laser field between the excited state
$\left\vert e\right\rangle $ and, respectively, the ground state
$\left\vert g\right\rangle $ and the spin up state $\left\vert
\uparrow \right\rangle $, form a $\Lambda$-type energy diagram,
which facilitates the use of the principle of electromagnetically
induced transparency (EIT) to determine the nature of pairing in the
interacting Fermi gas of two hyperfine spin states: $\left\vert
\uparrow\right\rangle $ and $\left\vert \downarrow\right\rangle $.
EIT \cite{EIT}, in which a probe laser field experiences (virtually)
no absorption but steep dispersion when operating around an atomic
transition frequency, has been at the forefront of many exciting
developments in the field of quantum optics \cite{Arimondo96}. Such
a phenomenon is based on quantum interference, which is absent in
measurement schemes such as in Ref. \cite{baym}, where lasers are
tuned far away from single-photon resonance. In the context of
ultracold atoms, an important example is the experimental
demonstration of dramatic reduction of light speed in the EIT medium
in the form of Bose condensate \cite{SlowVelocity}. This experiment
has led to a renewed interest in EIT, motivated primarily at the
prospect of the new possibilities that the slow speed and low
intensity light may add to nonlinear optics \cite{boyd02} and
quantum information processing \cite{QuantumInformation}. More
recently, EIT has been used to spectroscopically probe ultracold
Rydberg atoms \cite{rydberg}. In this work, we will show how EIT can
be exploited to reveal the nature of pairing in
Fermi gases.%

Before we present our detailed calculation, let us first compare the proposed
EIT method with the RF spectroscopy method which is widely used in probing
Fermi gases nowadays.
In the latter \cite{chin04,torma04}, an atomic sample is prepared and an RF
pulse is applied to the sample which couples one of the pairing states to a
third atomic level $|3\rangle$. This is followed by a destructive measurement
of the transferred atom numbers using absorption laser imaging. The RF signal
is defined as the average rate change of the population in state $\left\vert
3\right\rangle $ during the RF pulse, which can be inferred from the measured
loss of atoms in $|\uparrow\rangle$. This process is repeated for another RF
pulse with a different frequency. In addition to sparking many theoretical
activities \cite{levin05,griffin05,baym06,strinati08,basu08}, this method has
recently been expanded into the imbalanced Fermi gas systems
\cite{schunck06,he08,stoof08,shin07,Schirotzek08}, where paring can result in
a number of interesting phenomena \cite{imbalanced}. A disadvantage of this
method is its inefficiency: The sample must be prepared repeatedly for each RF
pulse. In addition, for the most commonly used fermionic atom species, i.e.,
$^{6}$Li, the state $|3\rangle$ interacts strongly with the pairing states due
to the fact that all three states involved has pairwise Feshbach resonances at
relatively close magnetic field strength. This leads to so-called final state
effect \cite{schunck08} which greatly complicates the interpretation of the RF spectrum.

In the EIT method, by contrast, one can directly measure the
absorption or transmission spectrum of the probe light. Applying a
frequency scan faster than the lifetime of the atomic sample to the
weak probe field, the whole spectrum can be recorded continuously in
a nearly non-destructive fashion to the atomic sample. Furthermore,
EIT signal results from quantum interference and is extremely
sensitive to the two-photon resonance condition. The width of the
EIT transparency window can be controlled by the coupling laser
intensity and be made narrower than $E_{F}$. As we will show below,
this property can be exploited to detect the onset of pairing as the
pairing interaction shifts and destroys the two-photon resonance
condition. In addition, due to different selection rules compared
with the RF method, one can pick a different final state whose
interaction with the pairing states are negligible [see
Fig.~\ref{Fig:schematics}(c)], hence avoiding the final state
effects.

The paper is organized as follows. In Sec.~\ref{Sec:model}, we described the
model under study and define the key quantity of the proposal --- the
absorption coefficient of the probe light. In Sec.~\ref{Sec:quasi}, we present
the expression of the probe absorption coefficient and construct a
quasiparticle picture that will become convenient to explain the features of
the spectrum. The results are presented in Sec.~\ref{Sec:results}, where
spectral features at different temperatues are explained. We also show that
how EIT spectrum can be used to detect the onset of pairing. A breif summary
is presented in Sec.~\ref{Sec:summary}. Finally, we provide an appendix in
Sec.~\ref{Sec:appendix} where the derivation of the EIT spectrum is provided.
In particular, we include in this derivation the pairing fluctuations in the
framework of the pseudogap theory \cite{levin05}.

\section{Model}

\label{Sec:model}

Let us now describe our model in more detail, beginning with the definition of
$\omega_{i}$ and $\Omega_{i}$ as the temporal and Rabi frequencies of the
probe ($i=p$) and coupling ($i=c$) laser field of plane waves copropagating
with an almost identical wavevector $\mathbf{k}_{L}$ (along $z$ direction).
The system to be considered is a homogeneous one with a total volume $V$, and
can thus be described by operators $\hat{a}_{\mathbf{k},i}$ $(\hat
{a}_{\mathbf{k},i}^{\dag})$ for annihilating (creating) a fermionic atom in
state $\left\vert i\right\rangle $ with momentum $\hbar\mathbf{k}$, and
kinetic energy $\epsilon_{k}=\hbar^{2}k^{2}/2m$, where $m$ is the atomic mass.
Here, $\hat{a}_{\mathbf{k},i}$ are defined in an interaction picture in which
$\hat{a}_{\mathbf{k},e}$ $=\hat{a}_{\mathbf{k},e}^{\prime}e^{-i\omega_{p}%
t},\hat{a}_{\mathbf{k},g}$ $=\hat{a}_{\mathbf{k},g}^{\prime}e^{i\left(
\omega_{c}-\omega_{p}\right)  t},$ and $\hat{a}_{\mathbf{k},\sigma}$ $=\hat
{a}_{\mathbf{k},\sigma}^{\prime}$ $(\sigma=\uparrow,\downarrow)$, where
$\hat{a}_{\mathbf{k},i}^{\prime}$ are the corresponding Schr\"{o}dinger
picture operators.

In a probe spectrum, the signal to be measured is the probe laser field, which
is modified by a polarization having the same mathematical form as the probe
field according to \cite{ling96}
\begin{equation}
\frac{\partial\Omega_{p}}{\partial z}+\frac{1}{c}\frac{\partial\Omega_{p}%
}{\partial t}=i\frac{\mu_{0}\omega_{p}cd_{e\uparrow}}{2}P_{p}\equiv
\alpha\Omega_{p}\,, \label{alpha}%
\end{equation}
where $P_{p}$ is the slowly varying amplitude of that polarization, $d_{ij}$
is the matrix element of the dipole moment operator between states $\left\vert
i\right\rangle $ and $\left\vert j\right\rangle $, and $\mu_{0}$ and $c$ are
the magnetic permeability and the speed of light in vacuum, respectively. The
parameter $\alpha$ in Eq.~(\ref{alpha}) represents the complex absorption
coefficient of the probe light \cite{ling96}. By performing an ensemble
average of atomic dipole moment, we can express $\alpha$ as
\begin{equation}
\alpha=i\frac{\alpha_{0}}{\Omega_{p}}\frac{1}{V}\sum_{\mathbf{k},\mathbf{q}%
}\left\langle \hat{a}_{\mathbf{q},\uparrow}^{\dag}\hat{a}_{\mathbf{k}%
+\mathbf{k}_{L},e}\right\rangle \,e^{i\left(  \mathbf{k}-\mathbf{q}\right)
\cdot\mathbf{r}}, \label{spe}%
\end{equation}
where $\alpha_{0}\equiv\mu_{0}\omega_{p}c\left\vert d_{e\uparrow}\right\vert
^{2}$. The real and imaginary part of $\alpha$ correspond to the probe
absorption and dispersion spectrum, respectively.

To determine the probe spectrum, we start from the grand canonical Hamiltonian
$\hat{H}=\sum_{\mathbf{k}}\left(  \mathcal{\hat{H}}_{1\mathbf{k}%
}+\mathcal{\hat{H}}_{2\mathbf{k}}+\mathcal{\hat{H}}_{3\mathbf{k}}\right)  $,
where
\begin{align*}
\mathcal{\hat{H}}_{1\mathbf{k}}  &  =\left(  \epsilon_{k}^{\prime}-\delta
_{p}\right)  \hat{a}_{\mathbf{k},e}^{\dag}\hat{a}_{\mathbf{k},e}+\left(
\epsilon_{k}^{\prime}-\delta\right)  \hat{a}_{\mathbf{k},g}^{\dag}\hat
{a}_{\mathbf{k},g}\,,\\
\mathcal{\hat{H}}_{2\mathbf{k}}  &  =-\frac{1}{2}(\Omega_{c}\hat
{a}_{\mathbf{k}+\mathbf{k}_{L},e}^{\dag}\hat{a}_{\mathbf{k},g}+\Omega_{p}%
\hat{a}_{\mathbf{k}+\mathbf{k}_{L},e}^{\dag}\hat{a}_{\mathbf{k},\uparrow
})-h.c\,,\\
\mathcal{\hat{H}}_{3\mathbf{k}}  &  =\sum_{\sigma}\epsilon_{k}^{\prime}\hat
{a}_{\mathbf{k},\sigma}^{\dag}\hat{a}_{\mathbf{k},\sigma}-(\Delta\hat
{a}_{\mathbf{k},\uparrow}^{\dag}\hat{a}_{-\mathbf{k},\downarrow}^{\dag
}+h.c)\,,
\end{align*}
describe the bare atomic energies of states $|e\rangle$ and $|g\rangle$, the
dipole interaction between atoms and laser fields, and the mean-field
Hamiltonian for the spin up and down subsystem, respectively. Here,
$\epsilon_{k}^{\prime}=\epsilon_{k}-\mu$ with $\mu$ being the chemical
potential, $\delta_{p}=\hbar\left(  \omega_{p}-\omega_{e\uparrow}\right)  $
and $\delta_{c}=\hbar\left(  \omega_{c}-\omega_{eg}\right)  $ are the
single-photon detunings, and $\delta=\delta_{p}-\delta_{c}$ is the two-photon
detuning with $\omega_{ij}$ being the atomic transition frequency from level
$\left\vert i\right\rangle $ to $\left\vert j\right\rangle $. In arriving at
$\mathcal{\hat{H}}_{3\mathbf{k}}$, in order for the main physics to be most
easily identified, we have expressed the collisions between atoms of opposite
spins in terms of the gap parameter $\Delta=-UV^{-1}\sum_{\mathbf{k}}%
\langle\hat{a}_{-\mathbf{k},\downarrow}\hat{a}_{\mathbf{k},\uparrow}\rangle$
under the assumption of BCS paring, where $U$ characterizes the interaction
between $|\uparrow\rangle$ and $|\downarrow\rangle$ which, in the calculation,
will be replaced in favor of the $s$-wave scattering length $a_{s}$ via the
regularization procedure:
\[
\frac{m}{4\pi\hbar^{2}a_{s}}=\frac{1}{U}+\frac{1}{V}\sum_{\mathbf{k}}\frac
{1}{2\epsilon_{k}}\,.
\]
A more complex model including the pseudo-gap physics \cite{levin05} will be
presented later in the paper. Finally, we note that the effect of the
collisions involving the final state $\left\vert g\right\rangle $ in the RF
spectrum has been a topic of much recent discussion
\cite{baym06,strinati08,basu08}. In our model, the spectra are not limited to
the RF regime, and this may provide us with more freedom to choose $\left\vert
g\right\rangle $ (and $\left\vert e\right\rangle )$ that minimizes the final
state effect. In what follows, for the sake of simplicity, we ignore the
collisions involving states $\left\vert g\right\rangle $ (and $\left\vert
e\right\rangle $). In practice, the effects of final state interaction can be
minimized by choosing the proper atomic species \cite{stewart08} or hyperfine
spin states \cite{final}. In the example shown in Fig.~\ref{Fig:schematics}%
(c), it is indeed expected that $|g\rangle$ does not interact strongly with
either of the pairing state.

\section{Quasiparticle picture}

\label{Sec:quasi}

The part of the Hamiltonian describing the pairing of the fermions can be
diagonalized using the standard Bogoliubov transformation:
\begin{align*}
\hat{a}_{\mathbf{k},\uparrow}  &  =u_{k}\hat{\alpha}_{\mathbf{k},\uparrow
}+v_{k}\hat{\alpha}_{-\mathbf{k},\downarrow}^{\dag}\,,\\
\hat{a}_{-\mathbf{k},\downarrow}^{\dag}  &  =-v_{k}\hat{\alpha}_{\mathbf{k}%
,\uparrow}+u_{k}\hat{\alpha}_{-\mathbf{k},\downarrow}^{\dag}\,,
\end{align*}
where $u_{k}=\sqrt{\left(  E_{k}+\epsilon_{k}^{\prime}\right)  /2E_{k}}%
,v_{k}=\sqrt{\left(  E_{k}-\epsilon_{k}^{\prime}\right)  /2E_{k}}$, and
$E_{k}=\sqrt{\epsilon_{\mathbf{k}}^{\prime2}+\Delta^{2}}$ is the quasiparticle
energy dispersion. Now we introduce two sets of quasiparticle states
$|\pm1_{\mathbf{k}}\rangle$, representing the electron and hole branches,
respectively. The corresponding field operators are defined as
\[
\hat{\alpha}_{\mathbf{k},+1}\equiv\hat{\alpha}_{\mathbf{k},\uparrow
}\,,\;\;\;\hat{\alpha}_{\mathbf{k},-1}\equiv\hat{\alpha}_{-\mathbf{k}%
,\downarrow}^{\dag}\,,
\]
in terms of which, the grand canonical Hamiltonian can be written as
\begin{widetext}
\begin{align}
\hat{H} &  =\sum_{\mathbf{k}}\left[  \left(  \epsilon_{k}^{\prime}-\delta
_{p}\right)  \hat{a}_{\mathbf{k},e}^{\dag}\hat{a}_{\mathbf{k},e}+\left(
\epsilon_{k}^{\prime}-\delta\right)  \hat{a}_{\mathbf{k},g}^{\dag}\hat
{a}_{\mathbf{k},g}+E_{k}\hat{\alpha}_{\mathbf{k},+1}^{\dag}\hat{\alpha
}_{\mathbf{k},+1}-E_{k}\hat{\alpha}_{\mathbf{k},-1}^{\dag}\hat{\alpha
}_{\mathbf{k},-1}\right.  \nonumber\\
&  \left.  -\left(  \frac{\Omega_{c}}{2}\hat{a}_{\mathbf{k}+\mathbf{k}_{L}%
,e}^{\dag}\hat{a}_{\mathbf{k},g}+h.c.\right)  -\left(  \frac{\Omega_{p}u_{k}%
}{2}\hat{a}_{\mathbf{k}+\mathbf{k}_{L},e}^{\dag}\hat{\alpha}_{\mathbf{k}%
,+1}+h.c\right)  -\left(  \frac{\Omega_{p}v_{k}}{2}\hat{a}_{\mathbf{k}%
+\mathbf{k}_{L},e}^{\dag}\hat{\alpha}_{\mathbf{k},-1}+h.c\right)  \right] \,.
\end{align}%
\end{widetext}A physical picture emerges from this Hamiltonian very nicely.
The state $\left\vert +1_{\mathbf{k}}\right\rangle $ ($\left\vert
-1_{\mathbf{k}}\right\rangle $) has an energy dispersion $+E_{k}\left(
-E_{k}\right)  $ and is coupled to the excited state $|e\rangle$ by an
effective Rabi frequency $\Omega_{p}u_{k}$ ($\Omega_{p}v_{k}$), which is now a
function of $k$. In the quasiparticle picture, our model becomes a double
$\Lambda$ system as illustrated in Fig.~\ref{Fig:schematics}(b). Let
$+\Lambda$ ($-\Lambda$) denote the $\Lambda$ configuration involving
$\left\vert +1_{\mathbf{k}}\right\rangle $ ($\left\vert -1_{\mathbf{k}%
}\right\rangle $). The $+\Lambda$ ($-\Lambda$) system is characterized with a
single-photon detuning of $\delta_{p}+E_{k}$ ($\delta_{p}-E_{k}$) and a
two-photon detuning of $\delta+E_{k}$ ($\delta-E_{k}$). In thermal equilibrium
at temperature $T$ (in the absence of the probe field), we have
\begin{equation}
\langle\hat{\alpha}_{\mathbf{k},+1}^{\dag}\hat{\alpha}_{\mathbf{k}^{\prime
},+1}\rangle=\delta_{\mathbf{k},\mathbf{k}^{\prime}}-\langle\hat{\alpha
}_{\mathbf{k},-1}^{\dag}\hat{\alpha}_{\mathbf{k}^{\prime},-1}\rangle
=\delta_{\mathbf{k},\mathbf{k}^{\prime}}f\left(  E_{k}\right)  ,\label{quasiparticle}%
\end{equation}
where
\begin{equation}
f\left(  \omega\right)  =\left[  \exp\left(  \omega/k_{B}T\right)  +1\right]
^{-1},
\end{equation}
is the standard Fermi-Dirac distribution for quasiparticles. Thus, as
temperature increases from zero, the probability of finding a quasiparticle in
state $\left\vert +1_{\mathbf{k}}\right\rangle $ increases while that in state
$\left\vert -1_{\mathbf{k}}\right\rangle $ decreases but the total probability
within each momentum group remains unchanged. Similarly, in the quasiparticle
picture, the probe spectrum receives contributions from two transitions%
\begin{align}
\alpha &  =i\frac{\alpha_{0}}{\Omega_{p}}\frac{1}{V}\sum_{\mathbf{k}%
,\mathbf{q}}e^{i\left(  \mathbf{k}-\mathbf{q}\right)  \cdot\mathbf{r}}%
\times\nonumber\\
&  \left[  u_{q}\rho_{e,+1}\left(  \mathbf{k}+\mathbf{k}_{L},\mathbf{q}%
\right)  +v_{q}\rho_{e,-1}\left(  \mathbf{k}+\mathbf{k}_{L},\mathbf{q}\right)
\right]  , \label{alpha+-}%
\end{align}
where $\rho_{i,\pm1}\left(  \mathbf{k},\mathbf{k}^{\prime}\right)  $
$=\left\langle \hat{\alpha}_{\mathbf{k}^{\prime},\pm1}^{\dag}\hat
{a}_{\mathbf{k},i}\right\rangle $ are the off-diagonal density matrix elements
in momentum space.


The equations for the density matrix elements can be obtained by
averaging, with respect to the thermal equilibrium defined in
Eq.~(\ref{quasiparticle}),  the corresponding Heisenberg's equations
of motion based upon Hamiltonian (3).  In the regime where the
linear response theory holds, the terms at the second order and
higher can be ignored, and the density matrix elements correct up to
the first order in $\Omega_{p}$ are then found to be
governed by the following coupled equations: %
\begin{align}
i\hbar\frac{d}{dt}\left[
\begin{array}
[c]{c}%
\rho_{e,\eta}\left(  \mathbf{k}+\mathbf{k}_{L},\mathbf{q}\right) \\
\rho_{g,\eta}\left(  \mathbf{k},\mathbf{q}\right)
\end{array}
\right]   &  =M_{\eta}\left[
\begin{array}
[c]{c}%
\rho_{e,\eta}\left(  \mathbf{k}+\mathbf{k}_{L},\mathbf{q}\right) \\
\rho_{g,\eta}\left(  \mathbf{k},\mathbf{q}\right)
\end{array}
\right] \nonumber\\
&  -\frac{\Omega_{p}}{2}\Lambda_{\eta}\left(  \mathbf{k}\right)
\delta_{\mathbf{k},\mathbf{q}},\;\left(  \eta=\pm1\right)  , \label{d rho/dt}%
\end{align}
where
\[
\Lambda_{+1}\left(  \mathbf{k}\right)  =\left(
\begin{array}
[c]{c}%
u_{k}f\left(  E_{k}\right) \\
0
\end{array}
\right)  ,\Lambda_{-1}\left(  \mathbf{k}\right)  =\left(
\begin{array}
[c]{c}%
v_{k}f\left(  -E_{k}\right) \\
0
\end{array}
\right)  ,
\]
and%
\[
M_{\eta}=\left[
\begin{array}
[c]{cc}%
\epsilon_{k}^{\prime}-\delta_{p}-\eta E_{k}-i\gamma & -\frac{\Omega_{c}}{2}\\
-\frac{\Omega_{c}^{\ast}}{2} & \epsilon_{k}^{\prime}-\delta-\eta E_{k}%
\end{array}
\right]  \,.
\]
Here we have introduced phenomenologically the parameter $\gamma$ which
represents the decay rate of the excited state $|e\rangle$. Inserting the
steady-state solution from Eq.~(\ref{d rho/dt}) into Eq.~(\ref{alpha+-}), we
immediately arrive at $\alpha\left(  \delta_{c},\delta\right)  =\alpha
_{+1}\left(  \delta_{c},\delta\right)  +\alpha_{-1}\left(  \delta_{c}%
,\delta\right)  $, where%
\begin{equation}
\alpha_{\pm1}\left(  \delta_{c},\delta\right)  =i\frac{\alpha_{0}}{2V}%
\sum_{\mathbf{k}}w_{\mathbf{k}}\left(  \delta_{c},\delta,\pm E_{k}\right)
f\left(  \pm E_{k}\right)  \left\{
\begin{array}
[c]{c}%
u_{k}^{2}\\
v_{k}^{2}%
\end{array}
\right.  , \label{wk}%
\end{equation}
with
\begin{equation}
w_{\mathbf{k}}\left(  \delta_{c},\delta,\omega\right)  =\frac{\epsilon
_{k}^{\prime}-\delta-\omega}{\lambda_{\mathbf{k}}\left(  \delta_{c}%
,\delta,\omega\right)  \left(  \epsilon_{k}^{\prime}-\delta-\omega\right)
-\left\vert \frac{\Omega_{c}}{2}\right\vert ^{2}}\,, \label{wk1}%
\end{equation}
and $\lambda_{\mathbf{k}}\left(  \delta_{c},\delta,\omega\right)
=\epsilon_{\mathbf{k}+\mathbf{k}_{L}}^{\prime}-\delta_{c}-\delta
-i\gamma-\omega$.%
\begin{figure}
[ptb]
\begin{center}
\includegraphics[
height=2.655in,
width=3.0469in
]%
{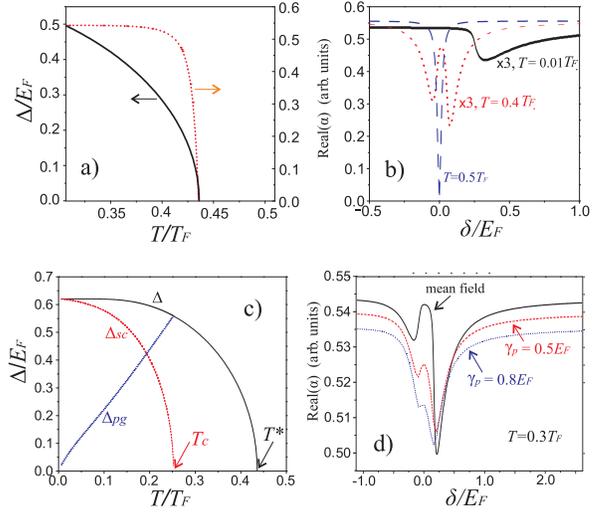}%
\caption{(Color online) (a) $\Delta$ (black solid curve) and the probe
absorption coefficient real$\left(  \alpha\right)  $ at $\delta=0$ (red dotted
curve) as functions of $T$, obtained from the mean-field BCS theory. (b)
Real$\left(  \alpha\right)  $ as a function of $\delta$ (absorption spectrum)
at different $T$. (c) $\Delta$, $\Delta_{sc}$ and $\Delta_{pg}$ as functions
of $T$ obtained from the pseudogap approach. ($\Delta_{sc}=0$ and
$\Delta=\Delta_{pg}$ when $T_{c}$ $<T<T^{\ast}$). (d) A comparison of various
absorption spectra at $T=0.3T_{F}$. The parameters are $\delta_{c}=0$,
$\gamma=380E_{F}$ ($\sim10$MHz), $\Omega_{c}=5E_{F}$ ($\sim0.1$MHz), and
$1/(k_{F}a_{s})=-0.1$.}%
\label{Fig:phaseDiagram}%
\end{center}
\end{figure}

\section{Results}

\label{Sec:results}

Examples of the probe absorption coefficient, Re($\alpha$), are presented in
Fig.~\ref{Fig:phaseDiagram}(a) and (b). For the results shown in this paper,
we choose $1/(k_{F} a_{s}) =-0.1$ where we denote $E_{F},k_{F}$, and
$T_{F}=E_{F}/k_{B}$ be Fermi energy, wavenumber, and temperature,
respectively, for the non-interacting Fermi gas. The black solid line in
Fig.~\ref{Fig:phaseDiagram}(a) represents the gap parameter in the mean-field
calculation, from which we can see that the critical temperature below which
the system exhibits pairing is given by $T_{c} = 0.435 T_{F}$ for the
parameters chosen. The dotted red curve in Fig.~\ref{Fig:phaseDiagram}(a)
represents the absorption coefficient at two-photon resonance $\delta=0$. We
can see that it remains at zero for $T>T_{c}$ but increases sharply once the
temperature drops below $T_{c}$. We note that this feature can be used as a
sensitive gauge for detecting the onset of Fermi pairing. With this being
emphasized, we now turn to explain the main spectroscopic features displayed
in Fig.~\ref{Fig:phaseDiagram}(b).

First, as long as $T>T_{c}$ where $\Delta=0$, one can show that the spectrum
is essentially independent of $T$ and
\[
\mathrm{Re}(\alpha) \propto\frac{\delta^{2}}{[(\delta+\delta_{c})
\delta-|\Omega_{c}/2|^{2}]^{2} + \delta^{2} \gamma^{2}}\,.
\]
From this expression, one can easily see that there exists around $\delta=0$ a
narrow transparency window with a width determined by the optical pumping rate
$\Gamma_{op}=\left\vert \Omega_{c}\right\vert ^{2}\gamma/\left[  4\left(
\delta_{c}^{2}+\gamma^{2}\right)  \right]  $ [see the blue dashed curve for
$T=0.5T_{F}$ in Fig.~\ref{Fig:phaseDiagram}(b)]. This feature can be most
easily understood from the bare state picture [Fig. \ref{Fig:schematics}(a)],
where state $\left\vert \uparrow\right\rangle $ is decoupled from state
$\left\vert \downarrow\right\rangle $ so that the spectrum is of EIT type for
a $\Lambda$ system involving $\left\vert e\right\rangle $, $\left\vert
g\right\rangle $, and $\left\vert \uparrow\right\rangle $. Further, because
states $\left\vert g\right\rangle $ and $\left\vert \uparrow\right\rangle $
share the same energy dispersion $\epsilon_{k}^{\prime}$, the two-photon
resonance condition $\delta=0$ holds for atoms of any velocity groups; the
absence of absorption at $\delta=0$ signals the existence of a coherent
population trapping state.

As $T$ decreases below $T_{c}$, a double-peak structure develops [see the red
dotted line for $T=0.4T_{F}$ in Fig.~\ref{Fig:phaseDiagram}(b)]. The two peaks
can be understood as contributed by the quasiparticle state $|+1_{\mathbf{k}%
}\rangle$ and $|-1_{\mathbf{k}}\rangle$, respectively. In the limit where $T$
is far below $T_{c}$ [see the black solid line for $T=0.01T_{F}$ in
Fig.~\ref{Fig:phaseDiagram}(b)], $+\Lambda$ system has negligible contribution
to the probe spectrum because there exists virtually no quasiparticles in
state $\left\vert +1_{\mathbf{k}}\right\rangle $. Thus, the spectrum is solely
contributed by $-\Lambda$ system, resulting a single-peak structure. However,
unlike the situations above $T_{c}$, here while the dispersion of an atom in
state $\left\vert g\right\rangle $ continues to be $\epsilon_{k}^{\prime}$,
the dispersion of a dressed particle in state $\left\vert -1_{\mathbf{k}%
}\right\rangle $ is $-E_{k}$. As a result, the effective two-photon
resonance condition $\epsilon_{k}^{\prime}-\delta+E_{k}=0$ is now
momentum dependent. Aside from a shift, the transparency window
becomes inhomogeneously broadened with a linewidth in the order of
$E_{F}$. A consequence of the momentum-dependence of the two-photon
resonance condition is that, for any given probe laser frequency,
only atoms with the `right' momentum result in perfect destructive
quantum interference. Consequently, Re($\alpha$) can no longer be
zero for any probe frequency. This underlies the sharp increase of
the probe absorption at $\delta=0$ below $T_{c}$ as shown in
Fig.~\ref{Fig:phaseDiagram}(a).

We also want to emphasize that the spectrum shown in
Fig.~\ref{Fig:phaseDiagram}(b) can be obtained by scanning the probe
laser frequency over a range on the order of $E_{F}\sim0.1$MHz. We
may take typical spectral features of the Fermi gas to be
$\delta\omega\sim0.1E_{F} \sim 10$KHz. To resolve such features,
using the energy-time uncertainty relation, we can use a scan rate
of 10KHz/0.1ms, then the total scan time can be estimated to be
around $1$ ms. As this time is much shorter compared with the
typical lifetime of the Fermi gas, this method can be regarded as
nearly non-destructive. This demonstrates the great efficiency of
the EIT probe.

In a more realistic model where pair fluctuations are included, gap $\Delta$
is divided into a BCS gap $\Delta_{sc}$ for condensed (BCS) pairs below
$T_{c}$ and a pseudogap $\Delta_{pg}$ for preformed (finite momentum) pairs
below temperature $T^{\ast}$ according to $\Delta^{2}=$ $\Delta_{sc}%
^{2}+\Delta_{pg}^{2}$ \cite{levin05}. Results including pseudogap physics are
illustrated in Fig.~\ref{Fig:phaseDiagram}(c) and (d) and the detailed
derivations can be found in the Appendex. In contrast to the weakly
interacting regime, where $T^{\ast}$ is virtually the same as $T_{c}$,
$T^{\ast}$ is much higher than $T_{c}$ in strongly interacting regime as is
clearly the case of present study according to Fig.~\ref{Fig:phaseDiagram}(c).
It needs to be stressed that pair fluctuations can result in a finite lifetime
$\gamma_{p}^{-1}$ for preformed pairs which tend to broaden the spectral
features, so that only when $\gamma_{p}$ is sufficiently small can the
double-peak spectroscopic structure be resolved as Fig.~\ref{Fig:phaseDiagram}%
(d) demonstrates. Finally, the two-photon resonance here is only sensitive to
$\Delta$ because $E_{k}$ depends on the total gap $\Delta$ \cite{levin05}. As
a result, like its RF counterpart \cite{schunck06}, the EIT method cannot
distinguish between $\Delta_{sc}$ and $\Delta_{pg}$. However, the qualitative
features of Fig.~\ref{Fig:phaseDiagram}(a) are not changed as long as we
regard the corresponding critical temperature as $T^{\ast}$.

\section{Summary}

\label{Sec:summary}

In summary, we propose to use optical spectroscopy in an EIT setting
to probe the fermionic pairing in Fermi gases. We have demonstrated
that the EIT technique offers an extremely efficient probing method
and is capable of detecting the onset of pair formation (i.e.,
determining $T^{\ast}$) due to its spectral sensitivity. With a
sufficiently weak probe field, the whole spectrum may be obtained
with a nearly non-destructive fashion via a relatively fast scan of
probe frequency, without the need of repeatedly re-preparing the
sample. We note that in this work, we have focused on probing the
atomic system using photons. In the future, it will also be
interesting to study how we can use atomic Fermi gas to manipulate
the light. Superfluid fermions can serve a new type of nonlinear
media for photons. Finally, we want to remark that, in this work, as
a proof-of-principle, we have only considered a homogeneous system.
As usual, the trap inhomogeneity can be easily accounted for within
local density approximation. Nevertheless, we note that the
capability of detecting the onset of pairing remains the same even
in the presence of the trap. Furthermore, as optical fields are used
in this scheme, one may focus the probe laser beam such that only a
small localized portion of the atomic cloud is probed, hence there
is no need to average over the whole cloud.

\acknowledgments
We thank Randy Hulet for insightful discussions. This work is supported by the
US National Science Foundation (H.P., H.Y.L.), the US Army Research Office
(H.Y.L.), and the Robert A. Welch Foundation (Grant No. C-1669), and the W. M.
Keck Foundation (L.J., H.P.), and by the National Natural Science Foundation
of China under Grant No. 10588402, the National Basic Research Program of
China (973 Program) under Grant No. 2006CB921104, the Program of Shanghai
Subject Chief Scientist under Grant No. 08XD14017, the Program for Changjiang
Scholars and Innovative Research Team in University, Shanghai Leading Academic
Discipline Project under Grant No. B480 (W.Z.).

\section{Appendix: EIT Spectra Including Pseudogap}

\label{Sec:appendix}

In this appendix, we generalize the result of Eq. (\ref{spe}) for $\alpha$
valid under the mean-field BCS pairing to a more realistic situation where
pair fluctuations are included in the form of pseudogap. We show two different
ways to accomplish this generalization. The first is an approach used more
often by people working in the field of quantum optics. The second uses the
linear response theory \cite{mahan} more familiar in the field of condensed
matter physics.

\subsection{A Brief Account of Pseudogap Theory}

\label{sec:pseudogap}

First, let us highlight the results of pseudogap theory \cite{levin05} that
are relevant to our EIT spectrum calculation. When pairing fluctuations at
finite temperature are included in the framework of the pseudogap model
\cite{levin05}, the BCS gap equation and number equation are still valid.
However, the gap $\Delta$ is now regarded as the total gap divided into a BCS
gap $\Delta_{sc}$ for condensed (BCS) pairs below $T_{c}$ and a pseudogap
$\Delta_{pg}$ for preformed (finite momentum) pairs:
\[
\Delta^{2} = \Delta_{sc}^{2} + \Delta_{pg}^{2} \,.
\]
The onset of the total gap $\Delta$ occurs at temperature $T^{\ast}$, which is
greater than $T_{c}$. The system with preformed pairs is described by the
Green's function
\begin{equation}
G^{-1}(\mathbf{k},iw_{n})=G_{0}^{-1}(\mathbf{k},iw_{n})-\Sigma(\mathbf{k}%
,iw_{n}), \label{GreenFunction}%
\end{equation}
where the non-interacting Green's function
\begin{equation}
G_{0}^{-1}(\mathbf{k},iw_{n})=\left(  i\omega_{n}-\epsilon_{\mathbf{k}%
}^{\prime}\right)  ^{-1}, \label{G0}%
\end{equation}
and the self energy
\begin{align}
\Sigma(\mathbf{k},iw_{n})  &  =\Sigma_{sc}(\mathbf{k},iw_{n})+\Sigma
_{pg}(\mathbf{k},iw_{n})\nonumber\\
&  =\frac{\Delta_{sc}^{2}}{iw_{n}+\epsilon_{k}^{\prime}}+\frac{\Delta_{pg}%
^{2}}{iw_{n}+\epsilon_{k}^{\prime}+i\gamma_{p}}\,\,, \label{selfenergy}%
\end{align}
with $w_{n}$ being the fermi Matsubara frequency and $\gamma_{p}^{-1}$ the
finite lifetime of pseudogap pairs. The spectral function $A(\mathbf{k}%
,\omega)$ can be obtained from the Green's function via the relation
\[
A(\mathbf{k},\omega)=-2\operatorname{Im}G\left(  \mathbf{k},\omega
+i0^{+}\right)  ,
\]
which, with the help of Eqs.~(\ref{GreenFunction}), (\ref{G0}), and
(\ref{selfenergy}), is found to be given by
\begin{equation}
A(\mathbf{k},\omega)=\frac{2(\omega+\epsilon_{k}^{\prime})^{2}\gamma_{p}%
\Delta_{pg}^{2}}{[\omega^{2}-E_{k}^{2}]^{2}(\omega+\epsilon_{k}^{\prime}%
)^{2}+\gamma_{p}^{2}[\omega^{2}-E_{k}^{sc2}]^{2}}\,, \label{AL}%
\end{equation}
where $E_{k}^{sc}=\sqrt{{\epsilon_{k}^{\prime}}^{2}+\Delta_{sc}^{2}}$. In the
limit of $\gamma_{p}\rightarrow0$ and $E_{k}^{sc}\rightarrow{E_{k}}$, we
recover from Eq. (\ref{AL}) the spectral function under the BCS\ paring
\begin{equation}
A(\mathbf{k},w)=2\pi\lbrack u_{k}^{2}\delta(\omega-{E_{k}})+v_{k}^{2}%
\delta(\omega+{E_{k}})]\,. \label{A BCS}%
\end{equation}

\subsection{Quantum Optics Approach}

In order to develop a formalism which directly incorporates the spectral
function, we rewrite Eq. (\ref{wk}) in terms of the equal time correlation
function $h_{\mathbf{q},\mathbf{k}}\left(  t\right)  =\left\langle \hat
{a}_{\mathbf{q},\uparrow}^{\dag}\left(  t\right)  \hat{a}_{\mathbf{k}%
+\mathbf{k}_{L},e}\left(  t\right)  \right\rangle $ as%
\begin{equation}
\alpha=i\frac{\alpha_{0}}{\Omega_{p}}\frac{1}{V}\lim_{t\longrightarrow\infty
}\sum_{\mathbf{k},\mathbf{q}}h_{\mathbf{q},\mathbf{k}}\left(  t\right)
e^{i(\mathbf{k}-\mathbf{q})\cdot\mathbf{r}}, \label{alpha Heisenberg}%
\end{equation}
where the limit is introduced to indicate explicitly that we are interested in
the steady state spectrum. \ Here, $\hat{a}_{\mathbf{q},\uparrow}^{\dag
}\left(  t\right)  $ and $\hat{a}_{\mathbf{k}+\mathbf{k}_{L},e}\left(
t\right)  $ obey the Heisenberg equations of motion%
\begin{equation}
i\hbar\frac{d}{dt}\left(
\begin{array}
[c]{c}%
\hat{a}_{\mathbf{k+k}_{L},e}\\
\hat{a}_{\mathbf{k},g}%
\end{array}
\right)  =\hat{M}\left(
\begin{array}
[c]{c}%
\hat{a}_{\mathbf{k+k}_{L},e}\\
\hat{a}_{\mathbf{k},g}%
\end{array}
\right)  -\frac{\Omega_{p}}{2}\hat{a}_{\mathbf{k},\uparrow}\left(
\begin{array}
[c]{c}%
1\\
0
\end{array}
\right)  , \label{Heisenberg equations}%
\end{equation}
with
\begin{equation}
\hat{M}=\left[
\begin{array}
[c]{cc}%
\epsilon_{\mathbf{k}+\mathbf{k}_{L}}^{\prime}-\left(  \delta_{p}%
+i\gamma\right)  & -\frac{\Omega_{c}}{2}\\
-\frac{\Omega_{c}^{\ast}}{2} & \epsilon_{k}^{\prime}-\delta
\end{array}
\right]  . \label{M-Inverse}%
\end{equation}
Note that due to the dissipative nature of our model, strictly speaking, Eqs.
(\ref{Heisenberg equations}) should be those of quantum Langevin\ equations
containing the noise operators of the reservoir that gives rise to the decay
rate $\gamma$. Here, in anticipation that Eqs.~(\ref{Heisenberg equations})
will produce the right averages of our interest, we have ignored the noise
operators. \ We solve Eqs.~(\ref{Heisenberg equations}) for $\hat
{a}_{\mathbf{k+k}_{L},e}\left(  t\right)  $ in the limit of $t\rightarrow
\infty$ when the terms involving the initial operators have all died away, and
then combine it with $\hat{a}_{\mathbf{q},\uparrow}^{\dag}\left(  t\right)  $
to form
\begin{equation}
h_{\mathbf{q},\mathbf{k}}\left(  t\right)  =\frac{\Omega_{p}}{2}\int_{0}%
^{t}\left[  e^{-i\hat{M}\left(  t-t^{\prime}\right)  }\right]  _{11}%
G^{<}\left(  \mathbf{k},t^{\prime},t\right)  \delta_{\mathbf{k},\mathbf{q}%
}dt^{\prime}, \label{h}%
\end{equation}
where $[...]_{11}$ denotes the element at the first row and the first column
of the matrix inside the square bracket, and $G^{<}\left(  \mathbf{k}%
,t^{\prime},t\right)  =i\left\langle \hat{a}_{\mathbf{k},\uparrow}^{\dag
}\left(  t\right)  \hat{a}_{\mathbf{k},\uparrow}\left(  t^{\prime}\right)
\right\rangle $ is one of the Green's functions in real time. By substituting
$G^{<}\left(  \mathbf{k},t^{\prime},t\right)  $ in Eq.~(\ref{h}) with a
Fourier transformation of its counterpart in real frequency, $G^{<}\left(
\mathbf{k},\omega\right)  $, we are able to carry out the time integration in
Eq. (\ref{h}) explicitly, leading to
\[
h_{\mathbf{q},\mathbf{k}}\left(  t\rightarrow\infty\right)  =\delta
_{\mathbf{k},\mathbf{q}}\frac{\Omega_{p}}{2}\int_{-\infty}^{+\infty}%
\frac{d\omega}{2\pi}\left[  \frac{A\left(  \mathbf{k},\omega\right)  f\left(
\omega\right)  }{\hat{M}-\omega}\right]  _{11},
\]
where the use of a well-known relation: $G^{<}\left(  \mathbf{k}%
,\omega\right)  =if\left(  \omega\right)  A\left(  \mathbf{k},\omega\right)  $
\cite{mahan} has been made. Finally, replacing\ $[1/ ( \hat{M}-\omega) ]_{11}$
with $w_{\mathbf{k}}\left(  \delta_{c},\delta,\omega\right)  $, obtained with
the help of Eq. (\ref{M-Inverse}), we arrive at%

\begin{equation}
\alpha=i\frac{\alpha_{0}}{2V}\sum_{\mathbf{k}}\int_{-\infty}^{+\infty}%
\frac{d\omega}{2\pi}A\left(  \mathbf{k},\omega\right)  f\left(  \omega\right)
w_{\mathbf{k}}\left(  \delta_{c},\delta,\omega\right)  \label{final alpha}%
\end{equation}
where $w_{\mathbf{k}}\left(  \delta_{c},\delta,\omega\right)  $ is defined in
Eq.~(\ref{wk1}) of the main text. One can easily check that
Eq.~(\ref{final alpha}) reduces to Eq.~(\ref{wk}) in the limit of mean-field
BCS pairing when Eq.~(\ref{A BCS}) is used as the spectral function.

\subsection{Condensed Matter Approach}

In order to use the linear response theory widely used in condensed matter
physics, we first divide our system into a \textquotedblleft left
part\textquotedblright\ comprising two hyperfine spin states: $\left\vert
\uparrow\right\rangle $ and $\left\vert \downarrow\right\rangle $, whose
physics has been described in Sec.~\ref{sec:pseudogap}, a \textquotedblleft
right part\textquotedblright\ consisting of the coupling laser field and
states $\left\vert g\right\rangle $ and $\left\vert e\right\rangle $,
described by the Hamiltonian%
\begin{align*}
\hat{H}_{R}  &  =\underset{\mathbf{k}}{\sum}\left[  (\epsilon_{k}^{\prime
}-\delta_{p})\hat{a}_{\mathbf{k},e}^{\dag}\hat{a}_{\mathbf{k},e}+(\epsilon
_{k}^{\prime}-\delta)\hat{a}_{\mathbf{k},g}^{\dag}\hat{a}_{\mathbf{k}%
,g}\right] \\
&  -\left(  \frac{\Omega_{c}}{2}\underset{k}{\sum}\hat{a}_{\mathbf{k}%
+\mathbf{k}_{L},e}^{\dag}\hat{a}_{\mathbf{k},g}+h.c.\right)  ,
\end{align*}
and finally the coupling between the two parts induced by the probe field,
described by the tunneling Hamiltonian
\begin{align*}
\hat{H}_{T}  &  =-\frac{\Omega_{p}}{2}\underset{\mathbf{k}}{\sum}\hat
{a}_{\mathbf{k}+\mathbf{k}_{L},e}^{\dag}\hat{a}_{\mathbf{k},\uparrow}+h.c\\
&  \equiv\hat{A}+\hat{A}^{\dag}\text{.}%
\end{align*}
Next, we change $H_{R}$ into a diagonal form
\begin{equation}
H_{R}=\underset{\mathbf{k}}{\sum}\,\left[  E_{k}^{\alpha}\hat{\alpha
}_{\mathbf{k}}^{\dag}\hat{\alpha}_{\mathbf{k}}+E_{k}^{\beta}\hat{\beta
}_{\mathbf{k}}^{\dag}\hat{\beta}_{\mathbf{k}}\right]  \,,
\end{equation}
in terms of a pair of dressed state operators, $\hat{\alpha}$ and $\hat{\beta
}$, defined via the transformation
\begin{equation}
\left[
\begin{array}
[c]{c}%
\hat{a}_{\mathbf{k}+\mathbf{k}_{L},e}\\
\hat{a}_{\mathbf{k},g}%
\end{array}
\right]  =\left[
\begin{array}
[c]{cc}%
u_{k}^{\alpha} & u_{k}^{\beta}\\
v_{k}^{\alpha} & v_{k}^{\beta}%
\end{array}
\right]  \left[
\begin{array}
[c]{c}%
\hat{\alpha}_{\mathbf{k}}\\
\hat{\beta}_{\mathbf{k}}%
\end{array}
\right]  ,
\end{equation}
where
\begin{align}
(u_{k}^{\alpha,\beta})^{2}  &  =(v_{k}^{\beta,\alpha})^{2}=\frac{1}{2}\left(
1\pm\frac{\zeta_{k}-\eta_{k}}{\sqrt{(\zeta_{k}-\eta_{k})^{2}+|\Omega_{c}|^{2}}
}\right)  ,\label{uv ab}\\
E_{k}^{\alpha,\beta}  &  =\frac{1}{2}\left(  \zeta_{k}+\eta_{k}\pm\sqrt
{(\zeta_{k}-\eta_{k})^{2}+|\Omega_{c}|^{2}} \right)  , \label{Ek ab}%
\end{align}
with $\zeta_{k}=\epsilon_{k+k_{L}}^{\prime}-\delta_{p}$ and $\eta_{k}%
=\epsilon_{k}^{\prime}-\delta$. In terms of the dressed state operators,
$\hat{A}$ becomes
\begin{equation}
\hat{A}=-\frac{\Omega_{p}}{2}\underset{\mathbf{k}}{\sum}\,\left[
u_{k}^{\alpha}\hat{\alpha}_{\mathbf{k}}^{\dag}\hat{a}_{\mathbf{k},\uparrow
}+u_{k}^{\beta}\hat{\beta}_{\mathbf{k}}^{\dag}\hat{a}_{\mathbf{k},\uparrow
}\right]
\end{equation}
and is in a form to which the linear response theory \cite{mahan} is directly
applicable. \ Following the standard practice, we then find
\begin{align}
\langle\hat{A}\rangle &  =\frac{\Omega_{p}^{2}}{4}\sum_{\mathbf{k}}\sum
_{\eta=\alpha,\beta}\left(  u_{k}^{\eta}\right)  ^{2}{\displaystyle\int
\limits_{-\infty}^{+\infty}}\frac{d\omega_{L}}{2\pi}A_{L}(\mathbf{k}%
,\omega_{L})\nonumber\\
&  {\displaystyle\int\limits_{-\infty}^{+\infty}}\frac{d\omega_{R}}{2\pi}%
A_{R}^{\eta}(\mathbf{k},\omega_{R})\frac{f(\omega_{R})-f(\omega_{L})}%
{\omega_{R}-\omega_{L}+i0^{+}}\,. \label{A}%
\end{align}
In Eq. (\ref{A}), $A_{L}(\mathbf{k},\omega_{L})$ is same as $A(\mathbf{k}%
,\omega_{L})$ defined in Eq. (\ref{AL}), while $A_{R}^{\eta}(\mathbf{k}%
,\omega_{R})$ is given by $2\pi\delta\left(  \omega_{R}-E_{k}^{\eta}\right)  $
because the right part is in a normal state described by the Green's function
$G_{\eta}^{-1}\left(  \mathbf{k},iw_{n}\right)  =iw_{n}-E_{k}^{\eta}$.
\ Integrating over $\omega_{R}$, we change Eq. (\ref{A}) into
\begin{equation}
\langle\hat{A}\rangle=\frac{\Omega_{p}^{2}}{4}\sum_{\mathbf{k}}\sum
_{\eta=\alpha,\beta}\left(  u_{k}^{\eta}\right)  ^{2}{\displaystyle\int
\limits_{-\infty}^{+\infty}}\frac{d\omega}{2\pi}A(\mathbf{k},\omega
)\frac{f(E_{k}^{\eta})-f(\omega)}{E_{k}^{\eta}-\omega+i0^{+}}\,, \label{A2}%
\end{equation}
where the dummy variable $\omega_{L}$ has been changed into $\omega$. We now
include the effect of the decay of the excited state phenomenologically by
replacing $\delta_{p}$ with $\delta_{p}-i\gamma$. We see that $E_{k}^{\eta}$
now become imaginary which signals the inability of the dressed states to hold
populations. This along with the fact that the dressed states here are the
superpositions of the initially empty states provide us with the justification
to set $f(E_{k}^{\eta})=0$ in Eq.~(\ref{A2}). With these considerations, we
finally arrive at
\begin{equation}
\langle\hat{A}\rangle=-\frac{\Omega_{p}^{2}}{4}\sum_{\mathbf{k}}%
{\displaystyle\int\limits_{-\infty}^{+\infty}}\frac{d\omega}{2\pi}%
A(\mathbf{k},\omega)f(\omega)w_{\mathbf{k}}\left(  \delta_{c},\delta
,\omega\right)  \label{final A}%
\end{equation}
where the use of Eqs. (\ref{uv ab}) and (\ref{Ek ab}) is made. It is clear
from Eq. (\ref{final alpha}) that $\alpha$ is proportional to $i\langle
A\rangle$ in Eq. (\ref{final A}).



\end{document}